# Origin of the Electronic Structure in Single-Layer FeSe/SrTiO$_3$ Films

Defa Liu[1,3,#,*], Xianxin Wu[1,4,#], Fangsen Li[5], Yong Hu[1], Jianwei Huang[1], Yu Xu[1,2], Cong Li[1,2], Yunyi Zang[3], Junfeng He[1], Lin Zhao[1], Shaolong He[1], Chenjia Tang[5], Zhi Li[5], Lili Wang[5], Qingyan Wang[1], Guodong Liu[1], Zuyan Xu[6], Xu-Cun Ma[5], Qi-Kun Xue[5], Jiangping Hu[1,2,7,8] and X. J. Zhou[1,2,7,8,*]

[1]*Beijing National Laboratory for Condensed Matter Physics, Institute of Physics, Chinese Academy of Sciences, Beijing 100190, China*
[2]*University of Chinese Academy of Sciences, Beijing 100049, China*
[3]*Max Planck Institute of Microstructure Physics, Weinberg 2, Halle 06120, Germany*
[4]*Department of Physics, Pennsylvania State University, University Park, Pennsylvania 16802, USA*
[5]*State Key Lab of Low-Dimensional Quantum Physics, Department of Physics, Tsinghua University, Beijing 100084, China*
[6]*Technical Institute of Physics and Chemistry, Chinese Academy of Sciences, Beijing 100190, China*
[7]*Songshan Lake Materials Laboratory, Dongguan 523808, China*
[8]*Beijing Academy of Quantum Information Sciences, Beijing 100193, China*

[#]These authors contribute equally to this work
*Corresponding author: liudf2006@126.com, XJZhou@iphy.ac.cn

The accurate theoretical description of the underlying electronic structures is essential for understanding the superconducting mechanism of iron-based superconductors. Compared to bulk FeSe, the superconducting single-layer FeSe/SrTiO$_3$ films exhibit a distinct electronic structure consisting of only electron Fermi pockets, due to the formation of a new band gap at the Brillouin zone (BZ) corners and an indirect band gap between the BZ center and corners. Although intensive studies have been carried out, the origin of such a distinct electronic structure and its connection to bulk FeSe remain unclear. Here we report a systematic study on the temperature evolution of the electronic structure in single-layer FeSe/SrTiO$_3$ films by angle-resolved photoemission spectroscopy. A temperature-induced electronic phase transition was clearly observed at 200 K, above which the electronic structure of single-layer FeSe/SrTiO$_3$ films restored to that of bulk FeSe, characterized by the closing of the new band gap and the vanishing of the indirect band gap. Moreover, the interfacial charge transfer effect induced band shift of ~ 60 meV was determined for the first time. These observations not only show the first direct evidence that the electronic structure of single-layer FeSe/SrTiO$_3$ films originates from bulk FeSe through a combined effect of

**an electronic phase transition and an interfacial charge transfer, but also provide a quantitative basis for theoretical models in describing the electronic structure and understanding the superconducting mechanism in single-layer FeSe/SrTiO$_3$ films.**

The pairing mechanism in iron-based superconductors remain a major unsolved issue in condensed matter physics. The unconventional superconductivity in iron-based superconductors arises from entangled interactions that involve structural distortion, magnetic order and nematic order [1-3]. Such complexities bring difficulties in identifying the key ingredients to superconductivity. Among all the iron-based superconductors [2-6], single-layer FeSe films grown on SrTiO$_3$ substrate (named FeSe/STO hereafter) have many unique properties, such as the simple crystal structure (Fig. 1a) [7], the high superconducting transition temperature (Tc > 65 K) [7-11], simple Fermi surface topology (Fig. 1b) [8-10], as well as the absence of nematicity [7,8] and phase separation [7]. These have made single-layer FeSe/STO films as an ideal platform to study the superconducting mechanism of iron-based superconductors.

For bulk FeSe and other FeAs-based superconductors, the electronic structure in the normal state can be well described by the theoretical calculations (Fig. 1d), hosting multiple hole and electron Fermi pockets at the Brillouin zone (BZ) center (Γ point) and corners (M points), respectively [2,3,12,13]. The presence of the hole and electron Fermi pockets are believed to be essential for the electron pairing and superconductivity. In contrast, single-layer FeSe/STO films with high temperature superconductivity exhibit a distinct electronic structure consisting of only electron Fermi pockets at M points (Fig. 1b), providing new insights into the superconducting mechanism [14-21]. The hole band at Γ point sinks well below the Fermi level ($E_F$) (Fig. 1c), resulting in the absence of the hole Fermi pockets around Γ point (Fig. 1b). The top position of the hole band at Γ point lies lower than the bottom of the electron bands at M point, leading to the formation of an indirect band gap

(illustrated by the gray rectangle in Fig. 1c). Moreover, a new band gap is developed between the electron bands and the hole bands below $E_F$ at M point (illustrated by the double arrow line in Fig. 1c). The formations of the new band gap at M point and the indirect band gap between Γ and M points are the two critical features in superconducting single-layer FeSe/STO films, which cannot be derived from the bulk FeSe (Fig. 1d) through a rigid band shift. Similar electronic structures have also been observed in other FeSe-based superconductors, $A_xFe_{2-y}Se_2$ ($A$ = K, Cs, Rb, Tl and so on) [22-24] and $(Li_{1-x}Fe_x)OHFe_{1-y}Se$ [25,26]. Although intensive theoretical calculations have been carried out, such as considering different atomic structures of FeSe film on $SrTiO_3$ substrate [27], different magnetic structures [27-30], electric field effect [29], oxygen deficiency effect [31], additional Se layer effect [32] and self-energy correction effect [33], few can reproduce the band structure of single-layer FeSe/STO films, leaving its origin still elusive. In addition, the interfacial charge transfer effect from the $SrTiO_3$ substrate to FeSe layer was observed [34,35], which plays an important role in realizing high temperature superconductivity in single-layer FeSe/STO films [36]. However, the effect of the charge transfer on the band structure and the relation between single-layer FeSe/STO films and bulk FeSe also remain unclear.

Several theoretical scenarios have been proposed to understand the high Tc superconductivity in single-layer FeSe/STO films, including the electron-phonon interaction [14-17], nematic fluctuations [18], orbital fluctuations [19], spin fluctuations [20] and incipient band pairing [21]. To verify the pairing mechanism and identify the key ingredients to the high temperature superconductivity, it is critical to reveal the origin of the distinct electronic structure and the charge transfer effect on the band structure in single-layer

FeSe/STO films.

Here, we report a systematic study and theoretical understanding on the temperature evolution of the electronic structure in single-layer FeSe/STO films by angle-resolved photoemission spectroscopy (ARPES). The temperature-induced electronic phase transition was clearly observed at 200 K, above which the electronic structure of single-layer FeSe/STO films restored to that of bulk FeSe, characterized by the closing of the new band gap and the vanishing of the indirect band gap. Moreover, the interfacial charge transfer induced energy shift of ~ 60 meV is determined experimentally. These results reveal that the distinct electronic structure of the single-layer FeSe/STO films originates from the bulk FeSe through a combined effect of an electronic phase transition and an interfacial charge transfer effect. Our results provide a quantitative basis for theoretical models in understanding the electronic structure and superconducting mechanism in single-layer FeSe/STO films.

We first study the temperature dependence of the band structure at $\Gamma$ point as shown in Fig. 2. The hole band (labeled as $\alpha$ in Fig. 2a (i)) lies below $E_F$ at 20 K, as seen clearly in Fig. 2a (i). Upon increasing temperature (Fig. 2a (i-vi)), although the $\alpha$ band gradually loses its intensity at its top position, the overall spectral weight slightly moves up to $E_F$. To quantitatively track the band position, the energy distribution curves (EDCs) at $\Gamma$ point measured at different temperatures are plotted in Fig. 2b. Two peaks are clearly observed at 20 K. The broad peak ($\omega$ band in Fig. 2a (i)) [37] locates at a high binding energy of ~ 0.27 eV and the other sharp peak ($\alpha$ band in Fig. 2a (i)) lies closer to $E_F$ at a binding energy of ~ 0.08 eV. The broad $\omega$ peak can be well resolved at all temperatures; it shows little change with temperature as illustrated by the orange dashed line in Fig. 2b. On the other hand, due to the thermal broadening, the sharp $\alpha$ peak gradually disappears with increasing temperature and becomes nearly invisible above 180 K, as seen in Fig. 2b. In order to highlight the peak structure, we divide the EDCs at different temperatures by that at 300 K, and the $\alpha$ peak structure can then be resolved at all temperatures (red dashed lines in Fig. 2c). The peak positions are extracted by fitting the two peaks (see more details in Supplementary Materials) and the results are summarized in Fig. 2d. The $\omega$ band remains unchanged upon increasing temperature, while the $\alpha$ band gradually moves up from $\sim$ -80 meV at 20 K to $\sim$ -54 meV at 270 K.

Next, we turn to the temperature evolution of band structure at M point. As shown in Fig. 3a (i), a band gap is developed between the electron band (labeled as $\delta$) and hole band (labeled as $\gamma$) at 20 K. With increasing temperature, the $\delta$ band first remains stationary below 180 K (Fig. 3a (i - iii)), then starts to move down (Fig. 3a (iv)). The $\gamma$ band changes little in its position below 80 K (Fig. 3a (i, ii)), then starts to move up (Fig. 3a (iii- iv)). The relative position changes of the $\delta$ and $\gamma$ bands with temperature result in the decrease of the gap size; they touch each other at 200 K (Fig. 3a (v)) accompanied by the gap closing. Fig. 3b shows the original EDCs at M point where the two peaks at $\sim$ -0.1 eV and $\sim$ -0.05 eV originate from the $\gamma$ band top and $\delta$ band bottom, respectively. They get smeared out above 140 K. In order to highlight the peak evolution at high temperatures, Fig. 3c shows the EDCs at different temperatures divided by that at 300 K. It becomes clear that the two peaks at low temperatures gradually evolve into a single peak at high temperatures. Such a behavior can also be seen from the temperature dependence of the momentum distribution curves (MDCs) extracted at $E_F$ for $\delta$ band (Fig. 3d (i)) and at -0.115 eV for $\gamma$ band (Fig. 3d (ii)), respectively.

Below T = 200 K, the size of $2k_\delta$ remains unchanged (red solid square in Fig. 3e), while the size of $2k_\gamma$ gradually increases upon the elevation of temperature (red hollow square in Fig. 3e), in agreement with the EDCs analysis (Fig. 3c). With further elevation of the temperature, both $2k_\delta$ and $2k_\gamma$ increase indicating the downward shift of the δ band and the upward shift of the γ band above T = 200 K. The extracted band position (Fig. 3e) and the gap size at M point (Fig. 3f), as well as the indirect band gap size between Γ and M points (Fig. 3f), clearly indicate that an electronic phase transition occurs at T = 200 K. The closing of the new band gap and the vanishing of the indirect band gap above 200 K indicate the electronic structure of single-layer FeSe/STO films restores to that of bulk FeSe. A resistivity anomaly was also observed at ~ 200 K in single-layer FeSe/STO films [38] which may originate from the electronic phase transition. Furthermore, we cycled the sample temperature back to low temperature and both bands at Γ and M points are recovered (see more details in Supplementary Materials), confirming the intrinsic nature of the temperature-induced band evolution in Fig. 2 and Fig. 3.

To understand the observed phase transition, we carried out band structure calculations using the five-orbital tight-binding model (see more details in Supplementary Materials). It is generally believed that the structural parameters, electron correlation and the $SrTiO_3$ substrate are responsible for the distinct electronic structure in single-layer FeSe/STO films (Fig. 1c). As these can be renormalized into the hopping parameters [39], we simulated the band structure evolution of bulk FeSe (Fig. 4a) by tuning the nearest-neighbor intraorbital hopping $t_1$ and $t_2$ for the $d_{xz}$ and $d_{yz}$ orbitals (Fig. 4b). The variations of $t_1$ and $t_2$ are related to the change of the ratio between the Se height and the lattice constant (see more discussions in

Supplementary Materials). Fig. 4a (i) shows the calculated band structure of bulk FeSe in the normal state with two hole bands and two electron bands crossing $E_F$. A separation can be seen between the two degenerate points at M point as illustrated by the orange and magenta arrows in Fig. 4a (i). However, the existence of the band crossing around M point between the electron band with $d_{xy}$ orbital character (blue lines) and the hole band with $d_{yz}$ orbital character (green lines) indicates that this separation is different from the gap observed in single-layer FeSe/STO films. With the increase of $|t_1|$ and decrease of $|t_2|$, the two degenerate points move closer until they touch each other at M point (Fig. 4a (ii)), accompanied by their zero separation. Further increasing $|t_1|$ and decreasing $|t_2|$, the two degenerate points swap their positions (Fig. 4a (iii)), indicating that a topological phase transition occurs [39]. This process leads to the crossing of the electron band with $d_{xz}$ orbital character and the hole band with $d_{xy}$ orbital character around M point. As these two bands belong to the same representations [39], they hybridize to form a new band gap structure, as shown in Fig. 4a (iii). Moreover, the hole bands with $d_{xz}$ and $d_{yz}$ orbital characters at Γ point gradually moves down below $E_F$ (Fig. 4a (i - iii)) and eventually an indirect band gap is developed between Γ and M points (illustrated by the gray rectangle in Fig. 4a (iii)). Apparently, the calculated band structure in Fig. 4a (iii) captures the two critical experimental observations (Fig. 1c), the indirect band gap between Γ and M points and the new band gap at M point. The calculated band evolution (Fig. 4a (iii) to (i)) exhibits a qualitative correspondence to the experimental observations upon increasing temperature (Figs. 2 and 3).

We illustrate our experimental results in Fig. 4c and assign the orbital characters to each band based on the calculations in Fig. 4a. At low temperature (T = 20 K, Fig. 4c (i)), the new band gap at M point is caused by swapping the positions of the initial two degenerate points. It closes at T = 200 K (Fig. 3a (v) and Fig. 3f) where the two degenerate points touch each other. Although we cannot resolve the two degenerate points above 200 K (Fig. 3a (vi) and Fig. 3c) due to the thermal broadening, the continuous increase of both $2k_\delta$ and $2k_\gamma$ with increasing temperature above 200 K (Fig. 3e) indicates their positions have been swapped as illustrated in Fig. 4c (ii). At T = 300 K, the top of the hole band around $\Gamma$ point moves up to ~ -54 meV while the bottom of the electron band around M point moves down to ~ -72 meV, leading to the vanishing of the indirect band gap between $\Gamma$ and M points. If we assume the electron-hole compensation for the undoped single-layer FeSe/STO films, the electron doping-induced energy shift of ~ 60 meV is obtained (Fig. 4c (ii)). As the Se vacancies in FeSe layer was suggested to cause an effective hole doping [40], we attribute this electron doping to the interfacial charge transfer effect [34,35]. Strikingly, with the deduction of the charge transfer effect, the band structure of the single-layer FeSe/STO films at T = 300 K in Fig. 4c (ii) shows an excellent correspondence to that of the bulk FeSe in Fig. 4a (i) with the appearance of hole pockets at $\Gamma$ point (the dashed cyan line in Fig. 4c (ii) represents the $E_F$ for the undoped single-layer FeSe/STO films).

The observed temperature-induced band structure evolution shows an excellent correspondence to the theoretical calculations, demonstrating that the distinct electronic structure of the single-layer FeSe/STO films originates from the bulk FeSe through a combined effect of an electronic phase transition and an interfacial charge transfer. These results also have profound implications on understanding the physics and superconductivity of single-layer FeSe/STO films. It was proposed that the new band gap at M point in single-layer FeSe/STO films is caused by the spin-orbit coupling (SOC) [41]. However, the SOC induced gap is generally independent with temperature which is not consistent with our observations (see Fig. 3f). An alternative scenario was proposed that involves the orbital selective Mott physics [42,43]. In this picture, the $d_{xy}$ orbital is completely Mott localized while the other orbitals remain itinerant, manifested by the disappearance of the spectral weight for the $d_{xy}$ orbital band at high temperature [37,44,45]. Our results show the peak structure from the $\delta$ band with $d_{xy}$ orbital character at M point gradually smears out with increasing temperature (see the original EDCs in Fig. 3b), which is consistent with the previous studies [37,45]. However, the peak structure for the $\alpha$ band with $d_{xz}/d_{yz}$ orbital character at Γ point also exhibits a similar behavior (see the original EDCs in Fig. 2b). As these bands are located near $E_F$ and the peaks reemerge after we divide the EDCs (Figs. 2c and 3c), this behavior is likely caused by the thermal broadening. Our results ask for a reexamination of the orbital selective Mott physics in single-layer FeSe/STO films.

The nematicity breaks $C_4$ symmetry in iron-based superconductors, which can induce a strong splitting between $d_{xz}$ and $d_{yz}$ orbitals around M point [5,46]. It can be completely suppressed by electron doping [1]. The heavy charge transfer from the SrTiO$_3$ substrate to the single FeSe layer provides a natural interpretation for the absence of nematicity in single-layer FeSe/STO films. With the deduction of the charge transfer effect, the undoped single-layer FeSe/STO films at high temperature (T=300 K, Fig. 4c (ii)) behaves like a semimetal with the compensation of electrons and holes; while it behaves like a semiconductor at low temperature (T = 20 K, Fig. 4c (i)) with an indirect band gap, which is consistent with the initial calculations [27]. Upon electron doping, the single-layer FeSe/STO films first become metallic and then superconducting, which is consistent with the scanning tunneling microscopy/spectroscopy (STM/STS) study [36].

The simulated band evolution by tuning the nearest-neighbor intraorbital hopping $t_1$ and $t_2$ for $d_{xz}/d_{yz}$ orbital can successfully capture the critical experimental observations, such as the new band gap at M point, the indirect band gap between Γ and M points, and the temperature-induced phase transition. These suggest that the underlying mechanism, such as the structural parameters, electron correlation and the $SrTiO_3$ substrate, plays important roles in driving the observed phase transition in single-layer FeSe/STO films. Further elucidation of the exact underlying mechanism and its effect on the high temperature superconductivity ask for more theoretical investigations.

In summary, we have carried out the systematic study on the temperature evolution of the electronic structure in single-layer FeSe/STO films and observed the electronic phase transition at 200 K, characterized by the closing of the new band gap at M points and the vanishing of the indirect band gap between Γ and M points with increase of the temperature. The temperature-induced phase transition shows a consistent evolution picture with the band structure calculations. Moreover, the interfacial charge transfer induced energy shift of ~ 60 meV is also determined experimentally. The present results provide the first direct evidence that the electronic structure of the single-layer FeSe/STO films originates from the bulk FeSe through a combined effect of an electronic phase transition and an interfacial charge transfer effect. Our results provide new insights in understanding the electronic structure and high temperature superconductivity in single-layer FeSe/STO films.

**Acknowledgment** We thank financial support from the National Key Research and Development Program of China (Grant No. 2016YFA0300300, 2016YFA0300600, 2017YFA0302900 and 2018YFA0305600), the National Natural Science Foundation of China (Grant No. 11888101, 11922414 and 11874405), the Strategic Priority Research Program (B) of the Chinese Academy of Sciences (Grant No. XDB25000000 and XDB33010300), the Youth Innovation Promotion Association of CAS (Grant No. 2017013), and the Research Program of Beijing Academy of Quantum Information Sciences (Grant No. Y18G06).

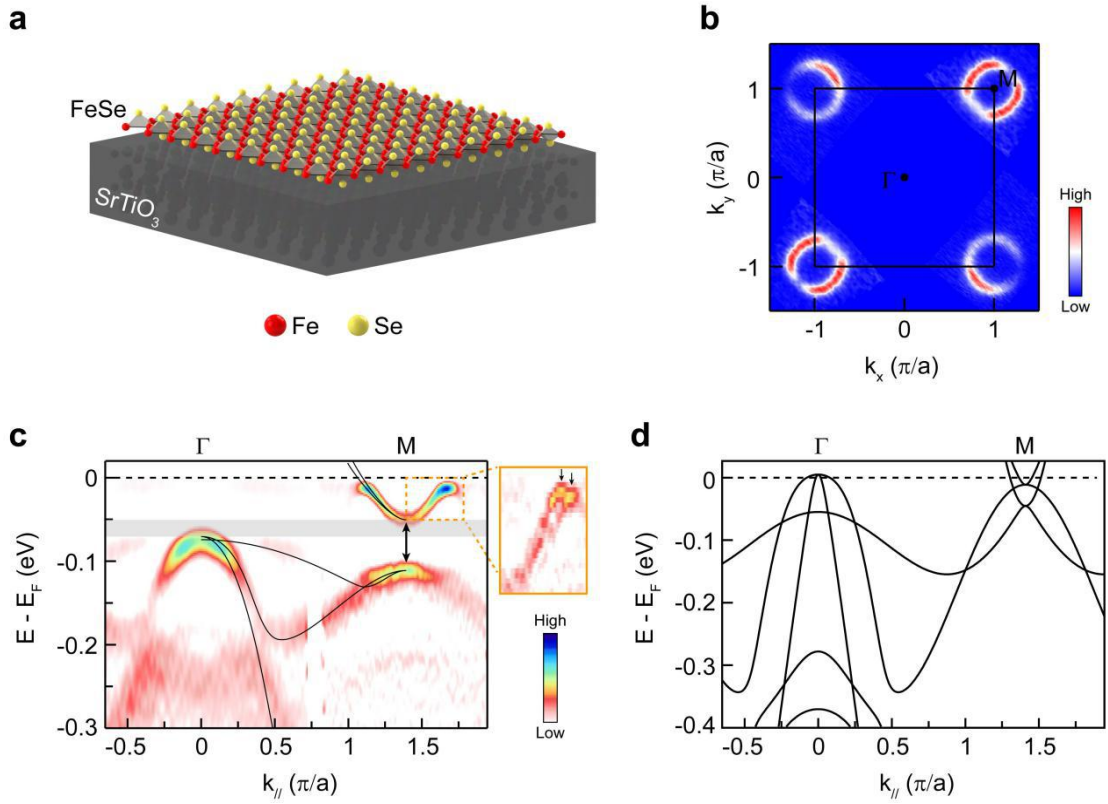

**Fig. 1. Electronic structure comparison of the superconducting single-layer FeSe/STO films to the bulk FeSe. a,** Illustration of single FeSe layer grown on SrTiO$_3$ substrate. **b,** Measured Fermi surface topology of the superconducting single-layer FeSe/STO films, consisting only electron Fermi pockets at M points. **c,** Measured band structure of single-layer FeSe/STO films along Γ-M direction. The overlaid black curves are the guidelines for the band structures. The gray rectangle illustrates the indirect band gap between Γ and M points. The double arrow line illustrates the new band gap developed at M point. The top-right zoom-in panel shows the two electron bands marked by the arrows. **d,** The calculated band structure for bulk FeSe in the normal state. Two hole bands and two electron bands cross $E_F$, which are different from the observed band structure of the single-layer FeSe/STO films in **c**.

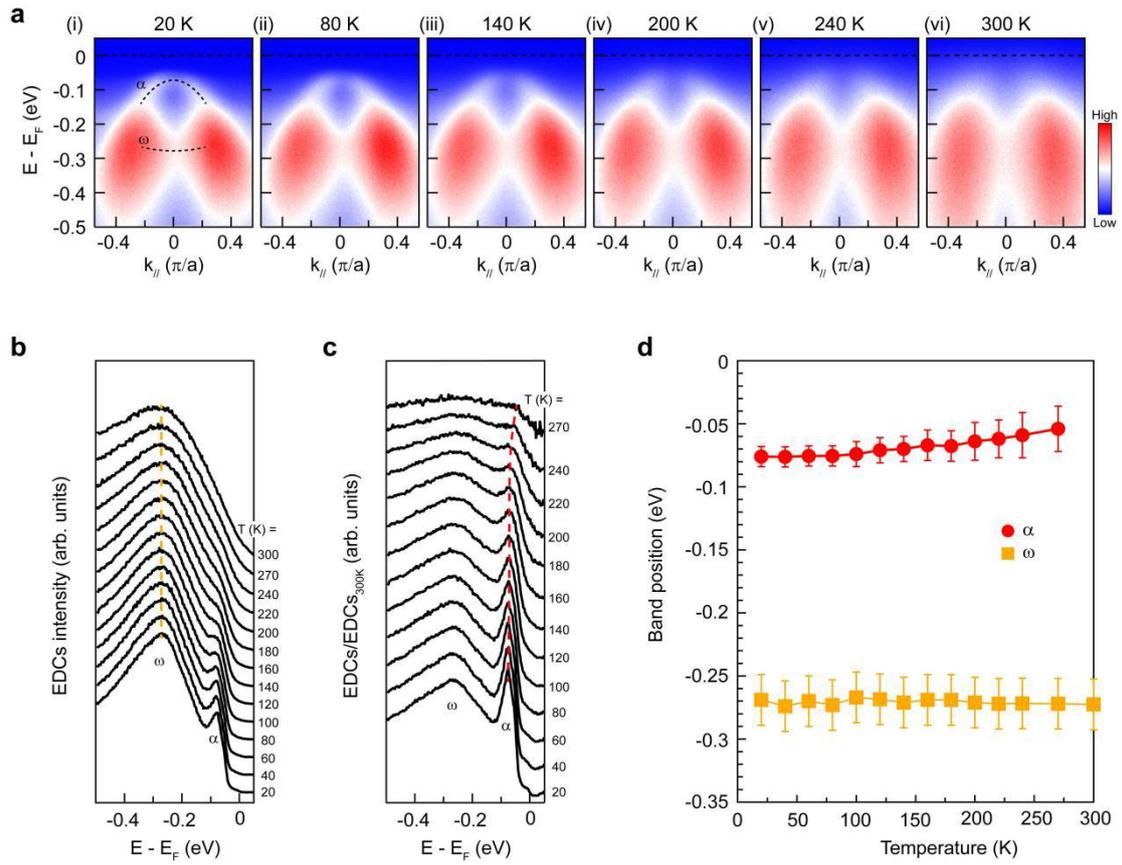

**Fig. 2. Temperature evolution of the band structure at Γ point. a,** The band structure at Γ point measured at different temperatures. **b,** The original EDCs at Γ point measured at different temperatures. The $\omega$ band shows little change with increasing temperature as illustrated by the dashed orange line. **c,** The divided EDCs at different temperatures obtained by dividing the original EDCs in **b** by the one at 300 K. The $\alpha$ band shows an upward movement with increasing temperature as illustrated by the dashed red line. **d,** Temperature evolution of the band positions for the $\alpha$ and $\omega$ bands.

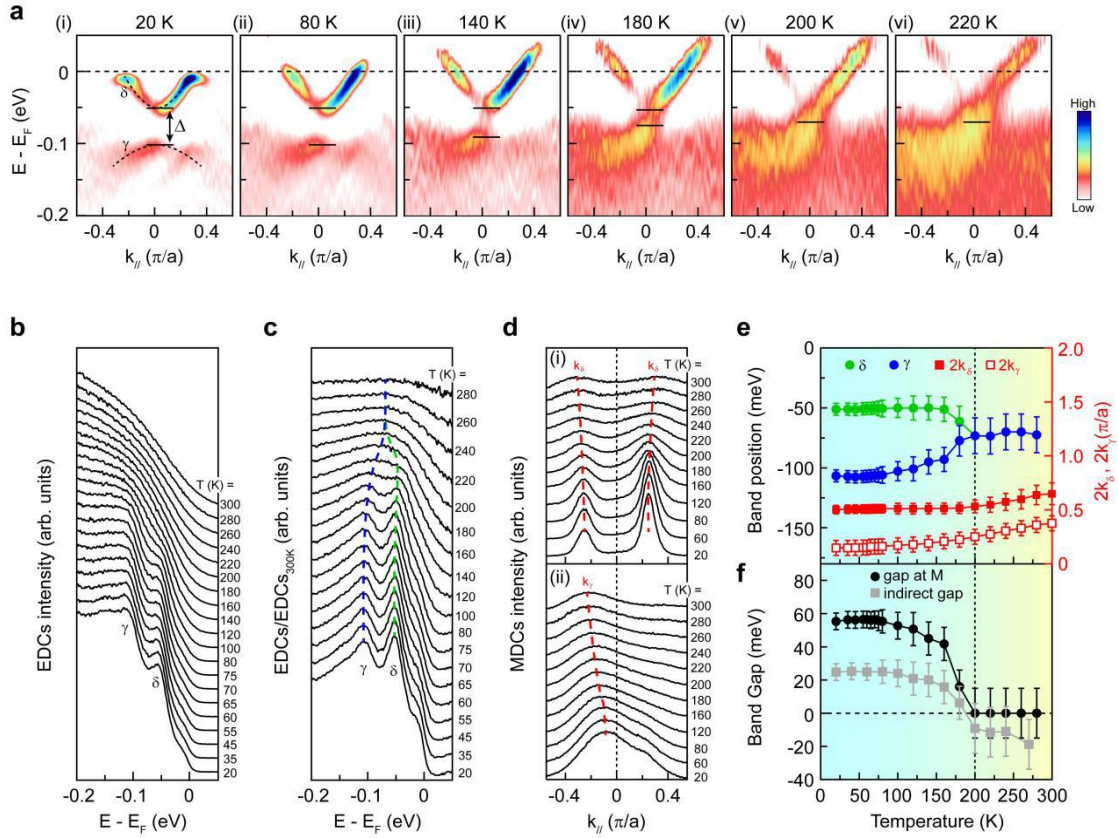

**Fig. 3. Temperature-induced phase transition at M point. a,** The temperature evolution of the band structure at M point. These images are second derivatives of the original data with respect to the energy. The Fermi-Dirac function is divided to highlight the unoccupied state above $E_F$. The electron band and hole band are labeled as $\delta$ and $\gamma$ bands, respectively, as shown in (i). The band gap ($\Delta$) gradually decreases with increasing temperature and eventually closes at 200 K. **b,** The original EDCs for different temperatures extracted at M point. **c,** The divided EDCs at different temperatures obtained by dividing the original EDCs in **b** with the one at 300 K. Two peaks gradually evolve into a single peak with increasing temperature, as illustrated by the dashed blue and green curves. **d,** Temperature dependence of the momentum distribution curves (MDCs) at $E_F$ for the $\delta$ band (i) and at -0.115 eV for $\gamma$ band (ii), respectively. The dashed red curves show the band positions at different temperatures. **e,** Temperature dependence of the extracted band positions and the size of $2k_\delta$ and $2k_\gamma$. **f,** Temperature dependence of the new band gap at M point and the indirect band gap between $\Gamma$ and M points. The gap at M point closes at 200 K suggesting a phase transition occurs at 200 K. The indirect band gap changes signs above 200 K indicating the vanishing of the indirect band gap characterized by the $\alpha$ band top at $\Gamma$ point lying higher than the $\delta$ band bottom at M point.

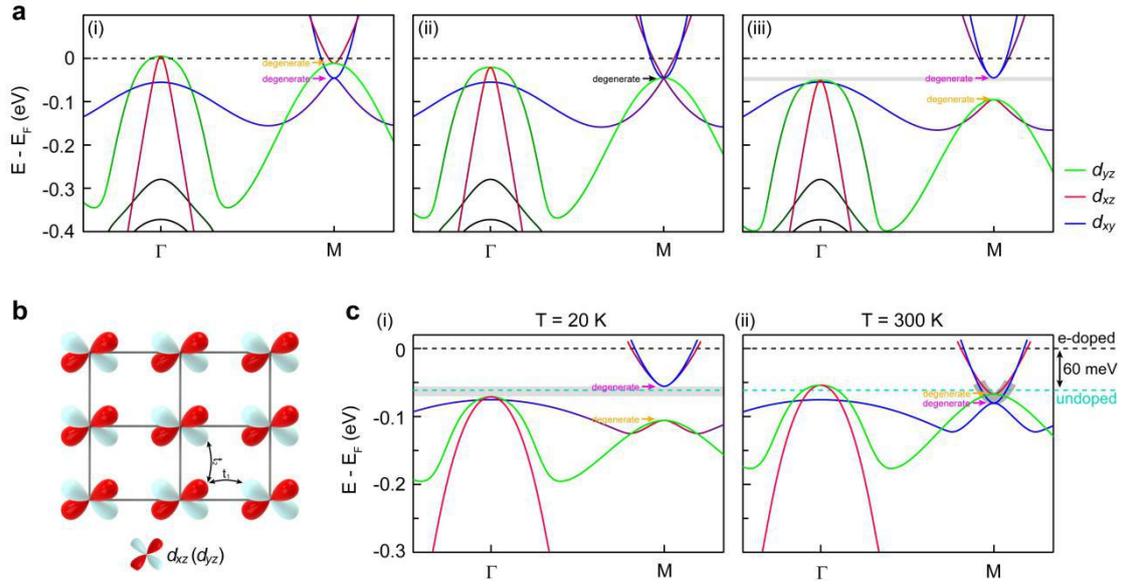

**Fig. 4. The origin of the electronic structure in single-layer FeSe/STO films. a,** Calculated band evolution by tuning the nearest-neighbor intraorbital hopping parameters $t_1$ and $t_2$ for $d_{xz}/d_{yz}$ orbital. $t_1$ = -0.1314 eV, $t_2$ = -0.4079 eV in (i) for the bulk FeSe in the normal state; $t_1$ = -0.1464 eV, $t_2$ = -0.4059 eV in (ii) for the phase transition point; $t_1$ = -0.1664 eV, $t_2$ = -0.4009 eV in (iii) for the single-layer FeSe/STO films. The band structure in (iii) captures two critical experimental observations: the new band gap at M point and the indirect band gap between Γ and M points. The degenerate points are labeled by the arrows. **b,** Schematic illustration for the nearest-neighbor intraorbital hopping for $d_{xz}/d_{yz}$ orbital. **c,** Extracted band structures of single-layer FeSe/STO films from the experimental results at low temperature (i) and high temperature (ii). With the deduction of the interfacial charge transfer effect, the $E_F$ shifts down of ~ 60 meV as illustrated by the dashed cyan lines for the undoped single-layer FeSe/STO films.